%% file: lpodio_printer_revised.tex
\documentclass[structabstract]{aa}  
\usepackage{graphicx}
\usepackage{txfonts}
\usepackage{natbib}
\def\farcs{\hbox{$.\!\!^{\prime\prime}$}}
\def\degr{\hbox{$^\circ$}}
\def\arcsec{\hbox{$^{\prime\prime}$}}
\newcommand{\niii}{[\ion{Ni}{ii}]}
\newcommand{\nai}{\ion{Na}{i}}
\newcommand{\feii}{[\ion{Fe}{ii}]}
\newcommand{\sii}{[\ion{S}{ii}]}
\newcommand{\caii}{[\ion{Ca}{ii}]}
\newcommand{\azi}{[\ion{N}{i}]}
\newcommand{\azii}{[\ion{N}{ii}]}
 
\newcommand{\oi}{[\ion{O}{i}]}

\newcommand{\Ha}{H$\alpha$}

\newcommand{\kms}{km\,s$^{-1}$}
\newcommand{\um}{$\mu$m}
\newcommand{\lam}{$\lambda$}

\newcommand{\cmc}{cm$^{-3}$}
\newcommand{\ang}{$\AA$}
\newcommand{\en}{n$_{e}$}
\newcommand{\nh}{n$_{H}$}
\newcommand{\te}{T$_{e}$}
\newcommand{\xe}{x$_{e}$}
\newcommand{\macc}{$\dot{M}_{\rm acc}$}
\newcommand{\mjet}{$\dot{M}_{\rm jet}$}
\newcommand{\pjet}{$\dot{P}_{\rm jet}$}
\newcommand{\poutflow}{$\dot{P}_{\rm outflow}$}
\newcommand{\rjet}{$r_{\rm jet}$}
\newcommand{\vjet}{$V_{\rm jet}$}
\newcommand{\vrad}{$V_{\rm rad}$}
\newcommand{\msolyr}{M$_{\odot}$\,yr$^{-1}$} 
\begin{document}
   \title{Tracing kinematical and physical asymmetries \\
     in the jet from DG Tau B}

   \author{Linda Podio
          \inst{1}
          \and
          Jochen Eisl\"offel\inst{2}
          \and
          Stanislav Melnikov\inst{3}
          \and
          Klaus W. Hodapp\inst{4}
          \and
          Francesca Bacciotti\inst{5}
          }

   \institute{Kapteyn Institute, 
          Landleven 12, 9747 AD Groningen, The Netherlands\\
              \email{podio@astro.rug.nl}
         \and
         Th\"uringer Landessternwarte Tautenburg, 
         Sternwarte 5, D-07778 Tautenburg, Germany\\
              \email{jochen@tls-tautenburg.de}
         \and
         Ulugh Beg Astronomical Institute, 
         Astronomical str. 33, 700052 Tashkent, Uzbekistan\\
             \email{stas@astrin.uzsci.net}
         \and
	 Institute for Astronomy, University of Hawaii, 
	 640 N. Aohoku Place, Hilo, HI 96720, USA\\
             \email{hodapp@ifa.hawaii.edu}
         \and
             INAF - Osservatorio Astrofisico di Arcetri, Largo Enrico Fermi 5,
             50125 Firenze, Italy\\
              \email{fran@arcetri.astro.it}
             }

   \date{Received ; accepted }


   \abstract
   {Stellar jets from young stars 
    can be highly asymmetric and have multiple velocity components.}
   {To clarify the origin of jet asymmetries and
     constrain the launch mechanism we study, as a test case, the physical and kinematical 
     structure of the prototypical asymmetric flow emitted by DG Tau B.}
   {The analysis of deep, high spectral resolution observations 
    taken at the KECK telescope 
     allows us to infer the properties and the spatial distribution of the 
     velocity components in the two jet lobes.
     From selected line ratios we derive the gas physical conditions 
     (the electron and total density, \en\, and \nh, 
     the ionisation fraction, \xe, and the temperature, \te), as a function
     of both distance from the source and gas velocity.
     The presence of dust grains in the jet is investigated by estimating the gas-phase 
     abundance of calcium with respect to its solar value.}
   {The detected lines show broad velocity profiles at the base of the jet (up to $\sim$100\,\kms)
     where up to three velocity components are detected. 
     At 5\arcsec\, from the source, however, only the denser and more excited high velocity components 
      survive and the lines are narrower ($\sim$10-30 \kms).
     The jet is strongly asymmetric both in velocity and in its physical structure. 
     The red lobe, which is slower ($\sim$140\,\kms)
     and more collimated (opening angle: $\alpha$$\sim$3-4\degr), presents low 
     ionisation fractions (\xe$\sim$0.1-0.4) and temperatures (\te$<$5\,10$^3$ K),
     while the total density is up to $\sim$2.5 10$^4$ \cmc.
     The blue lobe, faster ($\sim$-320\,\kms) and less collimated ($\alpha$$\sim$14\degr), 
     is also less dense (\nh$\sim$1 10$^4$ \cmc)
     but highly excited (\te\, up to $\sim$5 10$^4$ K and \xe\, up to 0.9).
     The estimated mass loss rate turns out to be similar in the two lobes 
     ($\sim$6-8 10$^{-9}$ \msolyr\,), while the flux of linear momentum
     is 3 times higher in the blue one ($\sim$2.5 10$^{-7}$ \msolyr\,\kms).
      Calcium is strongly depleted with respect to its solar abundance, 
     indicating that the jet contains dust grains. The depletion is lower for 
     higher velocities, consistent with dust destruction by shocks. }
   {The similar mass loss rate in the two lobes suggests that the ejection power is comparable 
     on the two sides of the system, as expected from a magneto-centrifugal ejection mechanism,  
     and that the observed asymmetries are due to different mass load and 
     propagation properties in an inhomogeneous environment. 
     The presence of dust grains implies that the jet is
     generated from a region of the disk extending beyond the dust sublimation radius.}

   \keywords{ISM: jets and outflows --
                Herbig-Haro objects --
                dust, extinction --
		Stars: formation
               }

   \titlerunning{Kinematical/physical asymmetries in the jet from DG Tau B}
   \authorrunning{L. Podio et al.}

   \maketitle
%

\section{Introduction}
\label{sect:intro}

Stellar jets from young stellar objects (YSO) play a key role in the star 
formation process and a wealth of information can be derived analysing their characteristic 
emission line spectra. 
The observed lines are collisionally excited in the shock waves generated by the 
interaction of the jet material with the interstellar medium or previously ejected 
matter and contain important information on the gas physics/kinematics. 
Different methods have been proposed to derive 
the physical conditions of the gas propagating in the jets from
emission line ratios, 
such as the comparison with the predictions of a
grid of shock models \citep[e.g., ][]{raga86,hartigan94}, or,
alternatively, spectral diagnostics techniques  
\citep[e.g., ][hereafter referred to as the BE technique]{bacciotti99}.

The application of these methods to analyse emission lines allows one to derive 
the jet structure on different scales, depending on the angular resolution of the observations:
from parsec scales to hundreds of AU
\citep{hartigan94,bacciotti99,podio06}, and
down to $\sim$15 AU from the emitting source with
space or Adaptive Optics-assisted observations
\citep{lavalleyfouquet00,bacciotti00,woitas02,hartigan07,melnikov09}. 
In particular, the analysis of emission lines can shed light on 
the gas conditions at the base of the flow and on the mechanism that generates it.
Proposed magneto hydro-dynamical (MHD) models suggest that the wind is
accelerated and collimated by 
magneto-centrifugal forces. It is not clear, however, where the jet originates: 
from the star itself \citep{sauty94}, from the radius at 
which the stellar magnetosphere truncates the disk (X wind, 
\citealt{shu00}), or from an extended region of the disk 
(Disk wind, \citealt{konigl00}).
 
In addition to the uncertainty on the launch mechanism, other
observed properties still lack a proper modeling. For example, 
velocity resolved observations of the jets from T Tauri stars 
and younger Class 0/I protostars show that these can 
be highly asymmetric and can have  different velocity components
\citep{hamann94,hirth94,hirth97,davis01,garcialopez08,garcialopez10}.
Despite some attempts to explain these observational features 
by means of the existing MHD models \citep[e.g., ][]{pesenti03,ferreira06} 
their origin is still unclear.
A viable possibility to clarify these aspects is to investigate the variation of the 
jet physical properties in asymmetric jets and in different velocity channels, 
as it has been attempted in recent works conducted at moderate and
high spatial/spectral resolution
\citep{lavalleyfouquet00,woitas02,coffey08,garcialopez08,garcialopez10,melnikov09,podio09}. 

Another interesting issue, which remains not fully clarified, is the dust content of jets. 
Refractory species such as Ca, Ni, Cr, Fe, are often locked onto dust grains, 
thus a strong depletion of their  gas phase abundance
is expected in the interstellar medium (ISM) \citep{savage96}. 
A few studies have investigated the gas phase abundance of Fe and Ni in HH jets 
\citep{beck-winchatz94,beck-winchatz96,mouri00,bohm01,nisini02} and
only recently the analysis has been extended to other refractory species such 
as Si, Ca, C, Cr, and Ti 
\citep{nisini05,nisini07,cabrit07,podio06,podio09,garcialopez08}.
These studies showed that the refractory  species may be depleted up to 90\%
in HH jets and molecular outflows, thus suggesting the presence of dust grains
in the ejected material.
This in turn  can put constraints on the region of the disk from where
jets originate.  In fact the stellar radiation destroys dust grains in the disk
up to the so called dust evaporation radius, R$_{evp}$.
This is located between 0.03 and 1.5 AU from the star in T Tauri
  and Herbig Ae/Be stars depending on the stellar
luminosity and the dust properties \citep{isella05,akeson05,eisner07},
but no direct observations and/or modeling of the inner disk structure 
are available for younger Class 0/I sources, still surrounded by circumstellar matter.
In general, we expect that  
a `dusty' jet can only arise from a region in the disk
extended beyond R$_{evp}$. 
The passage through a shock front, however,  can destroy part of the dust 
grains which may be transported in the jet, complicating the picture 
\citep[e.g., ][]{may00,guillet09}. Therefore, accurate studies of the
gas-phase abundance of refractory species are necessary to
determine the dust content in the jet and to constrain 
the shock efficiency in destroying the dust grains.

To improve our understanding of the physics of HH jets from 
YSO, particularly  their launch mechanism, 
we investigate the physical and kinematical
structure of the HH 159 jet emitted by the young source DG Tau B, focusing on 
the analysis of the velocity components detected at its base, the asymmetry
between the two jet lobes, and, finally, the estimate of the dust content in the
jet.  

HH 159 is a bipolar jet, first detected by \citet{mundt83}.  
The red lobe consists of a chain of bright knots extending 
to $\sim$55\arcsec\, from the source,
while the blue lobe, fainter and less collimated, is detected only up to $\sim$10\arcsec\, from
the source  \citep{mundt91,eisloffel98}. 
A large molecular outflow spatially coincident with the redshifted optical jet
has been detected in the CO lines by \citet{mitchell94,mitchell97}.
The driving star, DG Tau B, has been classified as a Class I
  source based on its
  spectral energy distribution \citep{watson04,luhman10}
 and at optical wavelengths it is obscured by circumstellar optically thick
  material detected in absorption through broadband imaging with the Hubble Space Telescope (HST)
\citep{stapelfeldt97}. 
The observed dark lane, elongated and perpendicular to the
  jet, indicates the presence of a circumstellar disk, confirmed 
 by $^{13}$CO millimeter observations \citep{padgett99,padgett00},
and flattened residual envelope.  
The source position has been determined from 3.5 cm VLA radio continuum
  observations by \citet{rodriguez95}, who locate it within the observed dark lane. 
Further HST studies in the near-infrared 
indicated the presence of a bipolar reflection nebula, 
whose axis of symmetry is coincident with the axis of the optical jet \citep{padgett99}.
The eastern lobe of the nebula, surrounding the blue jet lobe, is V-shaped,
suggesting that the nebula traces the walls of the blueshifted outflow
cavity. The western lobe is fainter and more collimated, and encompasses the
redshifted optical jet and the CO outflow.    

Previous spectroscopic \citep{mundt87,eisloffel98} and imaging 
\citep{mundt91,stapelfeldt97,padgett99} studies highlighted a strong asymmetry 
between the two lobes in morphology, velocity and degree of excitation. 
These studies, however, focused 
only on the \Ha\, and \sii\, emission 
lines as well as on the continuum emission in the optical and near-infrared bands.
On the contrary, in this paper we present for the first time very deep and high-spectral 
resolution spectroscopic observations of a large number of forbidden and 
permitted lines in the optical range from 5300~\ang\, to 8500~\ang.
The high spectral resolution coupled with long integration times highlight
a very complex velocity structure with multiple velocity components evolving 
along the jet. 
The large number of detected lines has allowed 
a detailed study of the gas physical conditions,
through the application of spectral diagnostic 
techniques to selected line ratios. This has  
provided a rich information on the excitation and dynamics of the
flow, as well as on the dust reprocessing.
The comparison of our results with those obtained for other jets 
provides means of investigating the nature of the detected velocity
components and the origin of asymmetric jets.      
This, in turn, can help understanding the generation of outflows 
and their role in young systems.


\section{Observations and data reduction}
\label{sect:obs}

   \begin{figure*}
     \centering
     \includegraphics[width=15.cm]{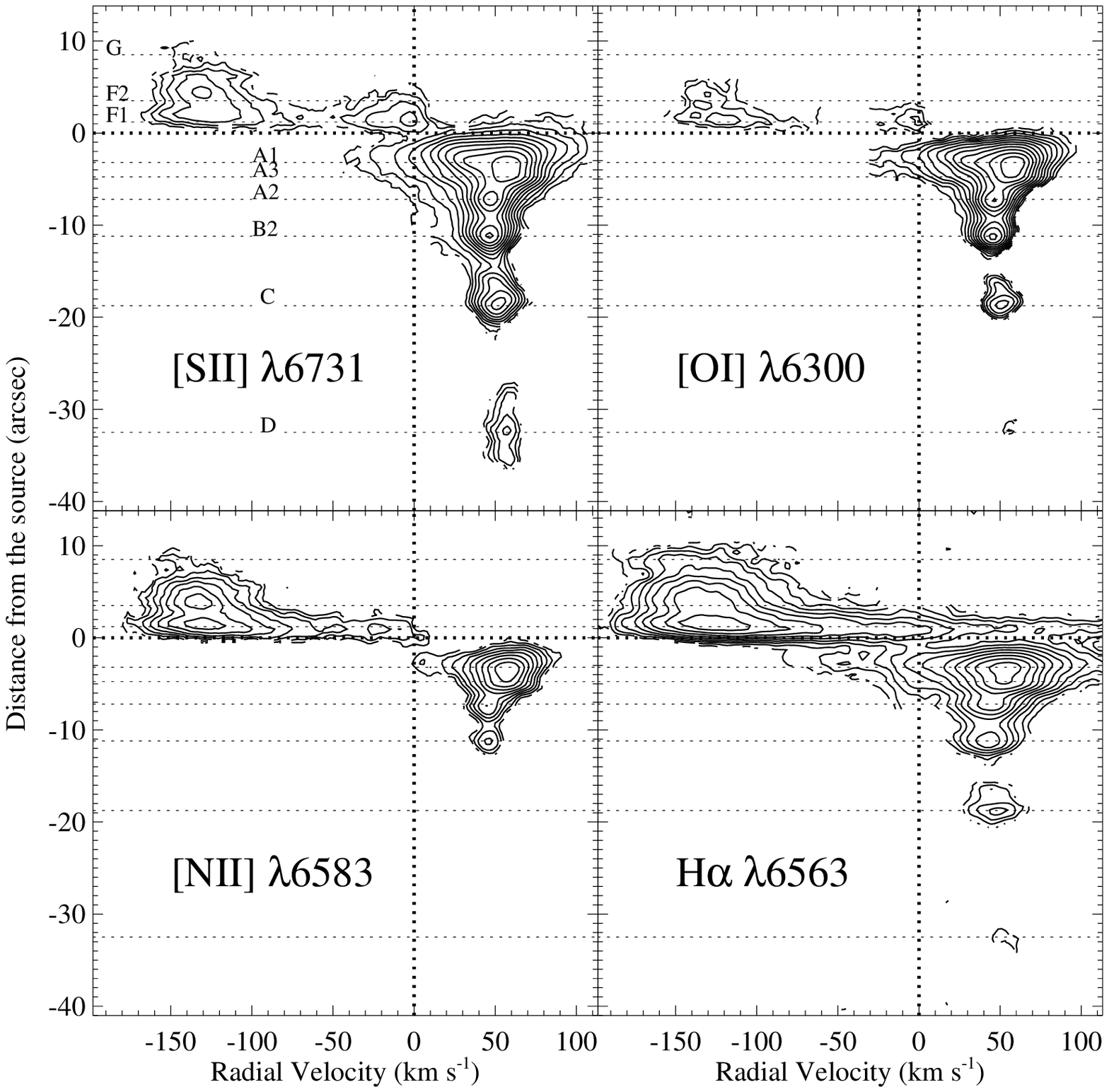}
   \caption{Position-Velocity diagrams of the jet from DG Tau B in the 
     strong optical emission lines \sii\lam6731, \oi\lam6300, \azii\lam6583, and 
     \Ha\lam6563  
   (angular resolution: FWHM$_{seeing}$$\sim$0.8\arcsec,
   spectral resolution: FWHM$_{inst}$$\sim$8.8~\kms).
     Contours are equally spaced in logarithmic scale. 
     The first contour is at 3$\sigma$ level for the \sii, \oi, and \azii\, lines 
   and at 5$\sigma$ for the \Ha\, line. The peak level is at 1103$\sigma$ for the \sii, 332$\sigma$ 
   for the \oi, 115$\sigma$  for the \azii, and 131$\sigma$ for the \Ha\, line.
   The vertical and horizontal dotted lines indicate the zero radial velocity with respect
   to the star and the spatial position of the source.
   The dashed horizontal lines indicate the position of the detected knots.}
   \label{fig:lines_pv}
    \end{figure*}
   \begin{figure*}
     \centering
     \includegraphics[width=15.cm]{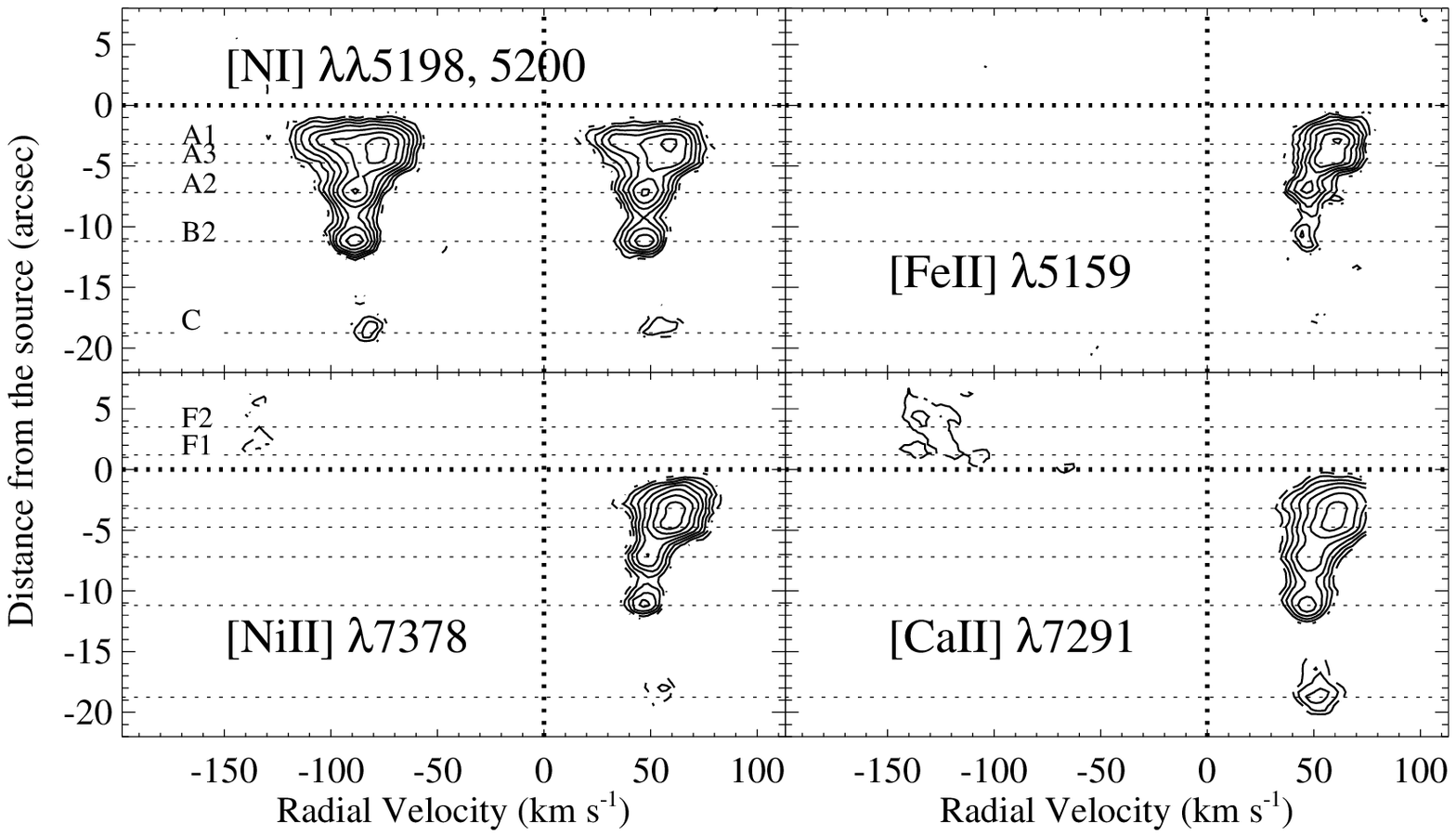}
   \caption{Position-Velocity diagrams of the jet from DG Tau B in the 
     \azi\lam\lam5198, 5200, \feii\lam5159, \niii\lam7378, \caii\lam7291 emission lines.
     Note that the velocity scale in the first panel is relative to the \azi\lam5200 line,
     while the scale for the \azi\lam\lam5198 plotted in the same
     panel is shifted by -136 \kms\, in velocity.
     Contours are equally spaced in logarithmic scale. 
     The first contour is at 2$\sigma$ level. The peak level is at 37$\sigma$ for the \azi, 
     24$\sigma$ for the \feii, 37$\sigma$  for the \niii, and 80$\sigma$ for the \caii\, line.
   The vertical and horizontal dotted lines indicate the zero radial velocity with respect
   to the star and the spatial position of the source.
   The dashed horizontal lines indicate the position of the detected knots.}
   \label{fig:faint_lines_pv}
    \end{figure*}

High resolution echelle spectra of  the jet from DG Tau B were obtained with
the HIRES spectrograph at the Keck I telescope \citep{vogt94}, located at the Keck
Observatory on Mauna Kea, Hawaii, on February 23/24, 1998. The
weather conditions were stable and photometric during the nights,  
providing a seeing of typically 0\farcs7-0\farcs8.
Four deep spectra of 2700\,s integration time were obtained using a long
slit decker of 28\arcsec$\times$1\farcs15, and a 2048$\times$2048 CCD Tektronix detector 
with 2$\times$binning in the spatial direction. This setting provided very high spectral resolution 
(R$\sim$34\,000, i.e. $\sim$8.8 \kms), spectral sampling of $\sim$0.036-0.052 \AA\,pixel$^{-1}$, 
(corresponding to $\sim$2.15 km\,s$^{-1}$), and a spatial scale of $\sim$0\farcs378 pixel$^{-1}$. 

The decker was aligned along the jet position angle (PA$\sim$295$^{\circ}$, \citealt{eisloffel98}). 
Since the decker is 28\arcsec\, long, i.e. shorter than the jet \citep{eisloffel98}, to cover the full 
jet length the observations were carried out by centering the decker on the jet source, 
and then shifting it along the jet PA towards the south-east and north-west relative 
to the source position, along the blue and the red lobe. In this way a length of $\sim$55\arcsec\, 
was covered along the spatial direction. The position angle tracking mode allowed us to keep the slit orientation at 
the chosen PA during the four exposures. 
By using the decker, 23 spectral orders overlapping along the spatial direction were obtained, covering a 
wavelength range from 5300~\ang\, to 8500~\ang. 

The raw frames were reduced following standard recipes with the
noao.imred.ccdred package in IRAF\footnote{IRAF is distributed by National Optical Astronomy Observatories, which is operated by the Association of Universities for Research in Astronomy, Inc. under contract with the National Science Foundation}, including overscan correction, bias and dark subtraction, flat-fielding, background subtraction, and wavelength calibration using Thorium-Argon arcs. The Laplacian Cosmic Ray Identification method (`lacos' method, \citealt{vandokkum01}) was used to clean the images from cosmics.

Because of the characteristic jet emission line spectrum we are interested in short sections of the echelle spectra. Thus, the latter were transformed into ordinary long-slit spectra centered around the detected lines. A flat-field frame obtained with a 28\arcsec\, long slit was used to calculate the tilt of the orders in the echelle images with respect to a normal long slit spectrum. Hence, the orders containing the emission lines were corrected for the estimated tilt. The same procedure was applied to the Thorium-Argon arcs frames. 

Wavelength calibration for the selected spectral orders was performed
by means of Thorium-Argon arcs, providing an accuracy of about 0.003~\AA. 
The observed velocities were corrected for the velocity of DG Tau
B with respect to the observer, hence the values of \vrad\, showed in
all the plots refer to the gas radial velocity with respect to the
driving source. 
Since the central source is obscured it is not
possible to estimate the source radial velocity from photospheric
absorption lines, e.g. Li\,6710~\AA. Thus, 
following \citet{eisloffel98} we assume that DG Tau B has the same 
heliocentric radial velocity as the surrounding molecular gas ($\sim$16.3 \kms, \citealt{mitchell94}),
which translates in a source velocity with respect to the observer of $\sim$46.4 \kms.

Finally, the four frames centered on different positions along the jet PA were aligned 
and then combined by estimating the position of the brightest 
emission knots.
Thus we obtain for each detected line a long slit spectrum covering $\sim$55\arcsec\, along the spatial
direction and $\sim$310 \kms\, along the spectral one (see the position-velocity (PV) diagrams in
Fig.~\ref{fig:lines_pv} and \ref{fig:faint_lines_pv} for selected
lines).
Note that in all the figures shown in this paper the distance from the source is positive along the blue
lobe and negative along the red one. 

At the end of the data reduction process, telluric features
overlapping with the jet emission lines 
and background caused by the overlap of adjacent orders were fitted by a polynomial surface and 
then subtracted. The extremely bright \oi\lam6300 telluric line could not be properly subtracted, 
thus the \oi\lam6300 line PV shows a $\sim$20\,\kms\, cut in the blue lobe centered at $\sim$-46.4 \kms 
(see Fig.~\ref{fig:lines_pv}). This cut is also clearly visible in the PV diagrams of the derived line ratios and 
of the jet physical properties (see Fig.~\ref{fig:ratios} and Fig.~\ref{fig:par}).

Note that the spectra are not flux calibrated, since no standard star observations are available.
Thus, the intensity of the lines in Fig.~\ref{fig:lines_pv} and \ref{fig:faint_lines_pv} is 
in arbitrary units with respect to the background noise, $\sigma$. 


\section{Results}
\label{sect:results}

   \begin{figure*}[!ht]
     \centering
     \includegraphics[width=6.3cm]{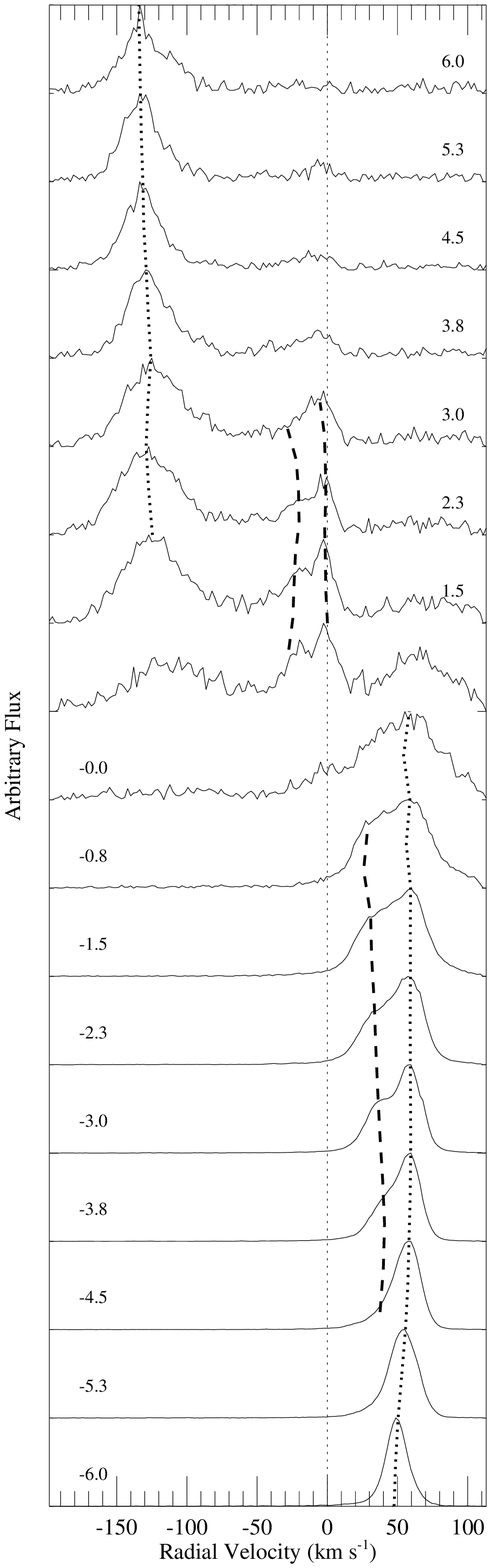}
     \includegraphics[width=6.3cm]{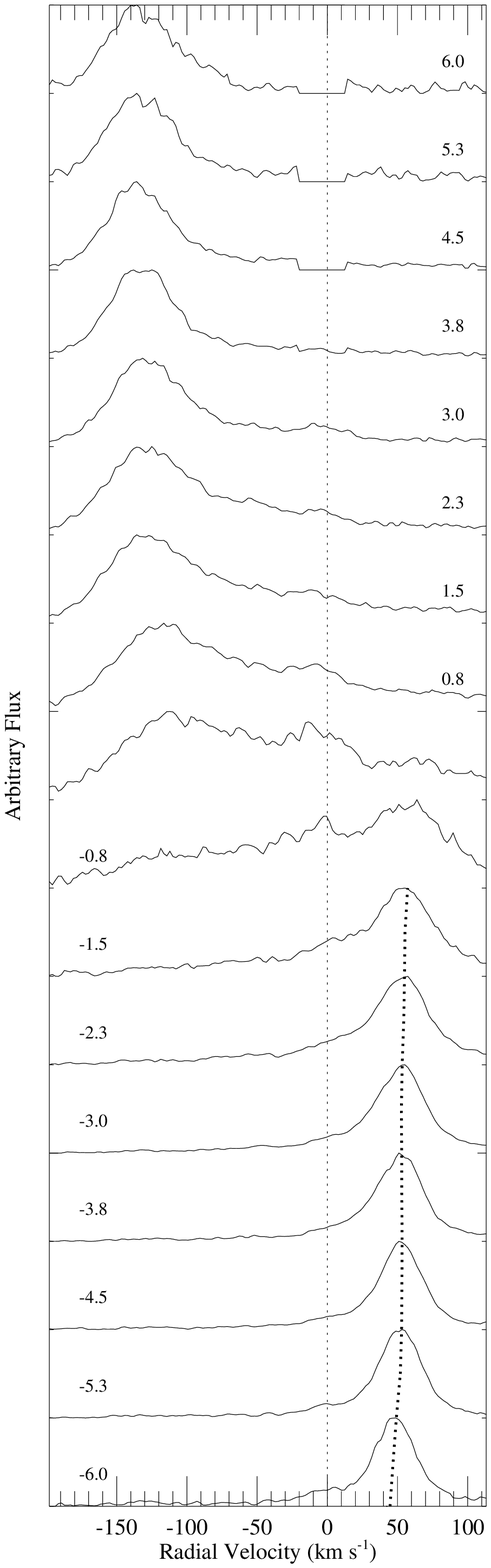}
   \caption{Normalized \sii\lam6731 ({\em left panel}) and \Ha\lam6563 ({\em right panel})
     velocity profiles along the first 6\arcsec\, of the blue and the red lobe (in arbitrary units), 
     obtained from the line PV diagrams in Fig.~\ref{fig:lines_pv} by spatially integrating the line spectral 
     profiles over the seeing FWHM ($\sim$0.8\arcsec, i.e. $\sim$2 pixels).
    The label at each  profile indicates  the distance from the star. 
   The vertical dotted line indicates the zero radial velocity with respect
   to the star.      
     The dotted and dashed thick lines indicate the velocity peak of the high 
     and  the low (and moderate) velocity components, respectively.
     Note that beyond $\sim$5\arcsec, only the high 
     velocity component survives in both lobes.}
   \label{fig:s31_profiles}
    \end{figure*}

The deep KECK/HIRES spectra (see Fig.~\ref{fig:lines_pv} and Fig.~\ref{fig:faint_lines_pv})
allowed us to analyse the emission of the HH 159 
jet up to $\sim$40\arcsec\, in the bright red lobe (knots A1, A3, A2, B2, C, 
and D, following the nomenclature by \citealt{eisloffel98}) 
and $\sim$10\arcsec\, in the fainter blue lobe (knot F1, F2,
and G).
Besides the \sii\lam\lam6716, 6731 and H$\alpha$\,\lam6563 lines, 
which have been previously observed by \citet{mundt87,mundt91} and \citet{eisloffel98}, 
we detected strong and broad emission in \oi\lam\lam6300, 6364, and in  
\azii\lam\lam6548, 6583 (see Fig.~\ref{fig:lines_pv}).
The jet is emitting also in a number of fainter and narrower emission lines, 
such as \feii\lam5159, \azi\lam\lam5198, 5200, \caii\lam7291, 
and \niii\lam7378, detected up to $\sim$20\arcsec\, (knot C) (see Fig.~\ref{fig:faint_lines_pv}). 
Finally, very faint emission in the \nai\, D lines at \lam\lam5890, 5896 is also 
detected at a $\sim$6$\sigma$ level in knots A1, A3.
Since the ionisation energy of \nai\, is very low, i.e. $\sim$5.1 eV, the 
\nai\, D lines are rarely observed in HH objects (see, e.g., HH 1 spectra by \citealt{solf88}). 
The detection of such lines suggests that the excitation conditions are quite 
low in the red lobe. 
This idea is confirmed by our analysis of the gas physical conditions 
(see Sect.~\ref{sect:jet_physics}).

\subsection{Jet kinematics: asymmetries and velocity components}
\label{sect:jet_kinematics}

   \begin{figure*}
     \centering
     \includegraphics[width=15.cm]{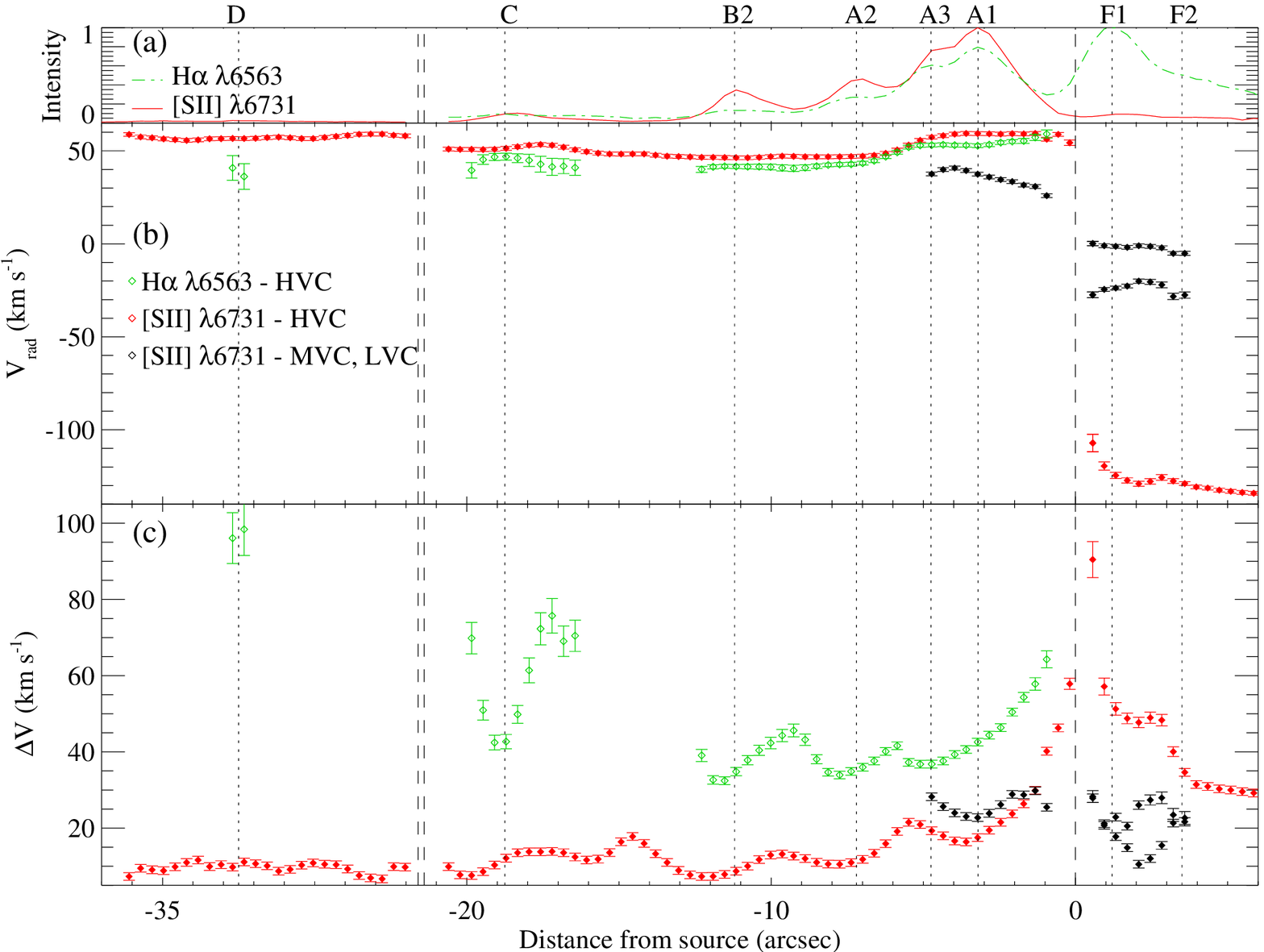}
   \caption{
     Jet kinematical properties as a function of the distance from the source.
     {\em From top to bottom:} 
     (a) intensity profiles of the \sii\, and the \Ha\, lines integrated over 
     their velocity profile and normalized to their peak intensity
     (red solid and green dotted-dashed line, respectively); 
     (b, c) peak radial velocity, \vrad, and full width at half maximum, $\Delta$V, of the different 
     velocity components in the jet as inferred from Gaussian deblending of the \sii\lam6731 
     (HVC: red points, MVC, LVC: black points) and the \Ha\,6563 (HVC:green points) line profiles 
     shown in Fig.~\ref{fig:s31_profiles}.} 
   \label{fig:kinematics}
    \end{figure*}

The PV diagrams of selected emission lines in Fig.~\ref{fig:lines_pv} show very broad 
(up to $\sim$200 \kms\, FWZI) and complex (up to 3 velocity components)
velocity profiles in the first few arcseconds from the source. 
Thanks to the high spectral resolution of the data ($\sim$8.8 \kms) we can resolve and
analyse each velocity component in the jet separately.
To derive the kinematical properties we extracted the jet spectral profile 
in the two lobes by spatially integrating the \sii\lam6731 and \Ha\, emission lines over the 
seeing full width at half maximum (FWHM$_{seeing}$$\sim$0\farcs7-0\farcs8, i.e. $\sim$2 pixels) 
at different distances along the jet length.
Fig.~\ref{fig:s31_profiles} 
show the variation of the \sii\lam6731 and \Ha\,  velocity profile in the red 
and the blue lobe up to $\sim$6\arcsec\, from the source.
The red lobe clearly shows the presence of a low
and a high velocity component (LVC and HVC), while in the blue lobe we 
detect three velocity components, 
at low, moderate, and high 
velocities (LVC, MVC, and HVC, respectively).
Interestingly, in both lobes, the low and moderate velocity components
fade away at $\sim$4-5\arcsec\, 
from the source, while the high velocity components survive further out.
This trend has been observed on a large sample of T Tauri stars but on 
much smaller scales, i.e between a few tens and 
200 AU from the source, by, e.g., \citet{hirth97} and \citet{pyo03}, 
and on similar or even larger scales at the base of HH jets from Class 0/I
sources \citep{garcialopez08,garcialopez10}.

By deblending and Gaussian fitting the spectral profiles shown in 
Fig.~\ref{fig:s31_profiles}
we obtain the gas radial velocity with respect to DG Tau B, \vrad, and 
the spectral width of each velocity component in the jet, $\Delta$V,
measured as the observed line full width at half maximum (FWHM$_{obs}$) 
corrected for the instrumental width (FWHM$_{inst}$$\sim$8.8 \kms).
This is done by assuming a Gaussian
profile for both the instrumental and the emission-line profile
(i.e. $\Delta$V = (FWHM$_{obs}$$^2$ - FWHM$_{inst}$$^2$)$^{1/2}$).
The error on the obtained determination of \vrad\, and $\Delta$V varies between 1-4 \kms.
The parameters derived by fitting the \sii\lam6731 and the \Ha\,\lam6563 line profiles
are plotted in Fig.~\ref{fig:kinematics} as a function of distance from the source.

The \sii\lam6731 line profiles in Fig.~\ref{fig:s31_profiles} show very
well defined velocity components.
In the red lobe the HVC radial velocity is decreasing in the first 14\arcsec\, from  the source
from $\sim$60\,\kms\, in the innermost knots A1, A3 down to $\sim$47~\kms\, in knots B2.
Then \vrad\, is increasing again moving towards the outer knots C (\vrad$\sim$50-55\,\kms\,)
and D (\vrad$\sim$55-60\,\kms\,).
This trend, clearly visible also in the line PV diagrams 
(see Fig.~\ref{fig:lines_pv} and \ref{fig:faint_lines_pv}),
 can be explained if the jet direction is changing due to
  precession or interaction with the ambient medium.
Observational evidences of precessing jets have been already reported 
by, e.g., \citet{giannini05}, \citet{hodapp05}, and \citet{caratti08}.
The LVC peak velocity instead varies between 25 and 42 \kms.
In the blue lobe much higher radial velocities are measured 
of $\sim$120-135 \kms\, in the 
HVC, $\sim$20-30 \kms\, in the MVC, and a few 
\kms\, in the LVC.
Given the jet inclination angle with respect to the line of sight of $\sim$65\degr\, estimated 
through proper motions measurements \citep{eisloffel98}, 
the bulk of the gas is moving with a velocity \vjet$\sim$140\,\kms\, in the red lobe, and
more than two times higher (\vjet$\sim $-320\,\kms) in the blue one.
These values are in agreement with those estimated by \citet{eisloffel98}.
The width, $\Delta$V,  of the high velocity component is decreasing with distance from
the source in both lobes. 
In the blue lobe we measure  $\Delta$V$\sim$60 \kms\, at $\sim$1\arcsec, decreasing to 
 $\sim$30 \kms\, at $\sim$6\arcsec. 
In the red lobe $\Delta$V decreases more steeply from $\sim$40-60 \kms\, within 1\arcsec\, 
from the source down to $\sim$15 \kms\, at $\sim$3\arcsec, and values between 6-20\,\kms\, for larger
distances.

The \Ha\,\lam6563 line profiles are much broader than the \sii\, profiles (FWZI up to $\sim$200 \kms) 
and the emission is very diffuse in this line, making it difficult to separate the different
velocity components.
Thus we only report the values of \vrad\, and $\Delta$V estimated for the HVC of the red lobe.
Fig.~\ref{fig:kinematics} shows that \vrad\, has the same trend  
as in the \sii\, lines, with only slightly lower values. 
The line width is much larger, though ($\Delta$V varies between $\sim$35 and $\sim$100 \kms).
A broader \Ha\, line profile was also noted by \citet{eisloffel98}.

The large velocity dispersion measured within a few arcseconds from the source and,
in particular, in the blue lobe, suggests that stronger shocks  
are present at the jet base, and in the blue lobe.
As we will show in the next section this reflects in higher excitation conditions in the blue lobe, 
that is the jet is highly asymmetric not only in
velocity but also in its physical structure.

\subsection{Jet physical structure: density and excitation}
\label{sect:jet_physics}

\begin{table*}
\caption[]{\label{tab:avg_par} Kinematical, physical, and dynamical properties of the jet from DG Tau B
in the high and the low velocity components (HVC and LVC, respectively) averaged over the first 5\arcsec\, 
of the red and the blue lobe.
The values of \vrad\, and $\Delta$V are derived from Gaussian fits to the \sii \lam6731 line profiles and
to the \Ha\,6563 (in parentheses).}
    \begin{tabular}[h]{cc|cc|cccc|cc}
      \hline
Lobe  & VC & \vrad & $\Delta$V   & n$_e$              & x$_e$ & T$_e$      & n$_H$              & \mjet             & \pjet        \\
      &    & \kms  & \kms      & 10$^{3}$ cm$^{-3}$ &       & 10$^{4}$ K & 10$^{4}$ cm$^{-3}$ & 10$^{-9}$ \msolyr & 10$^{-6}$ \msolyr \kms \\
\hline 
Red  & HVC & 58 (54) & 27 (46) & 3.3 & 0.22 & 0.45 & 1.49 & 4.2  & 0.6     \\
     & LVC & 35      & 26      & 1.9 & 0.17 & 0.40 & 1.29 & 2.2  & 0.2     \\
\hline  
Blue & HVC & -126    & 47      & 3.6 & 0.88 & 2.46 & 0.41 & 8.2  & 2.5     \\
     & LVC & -2      & 18      & 1.1 & 0.41 & 0.74 & 0.29 & 0.09 & 7~10$^{-4}$  \\
\hline 
      \end{tabular}
\end{table*}

   \begin{figure*}
     \centering
     \includegraphics[width=15.cm]{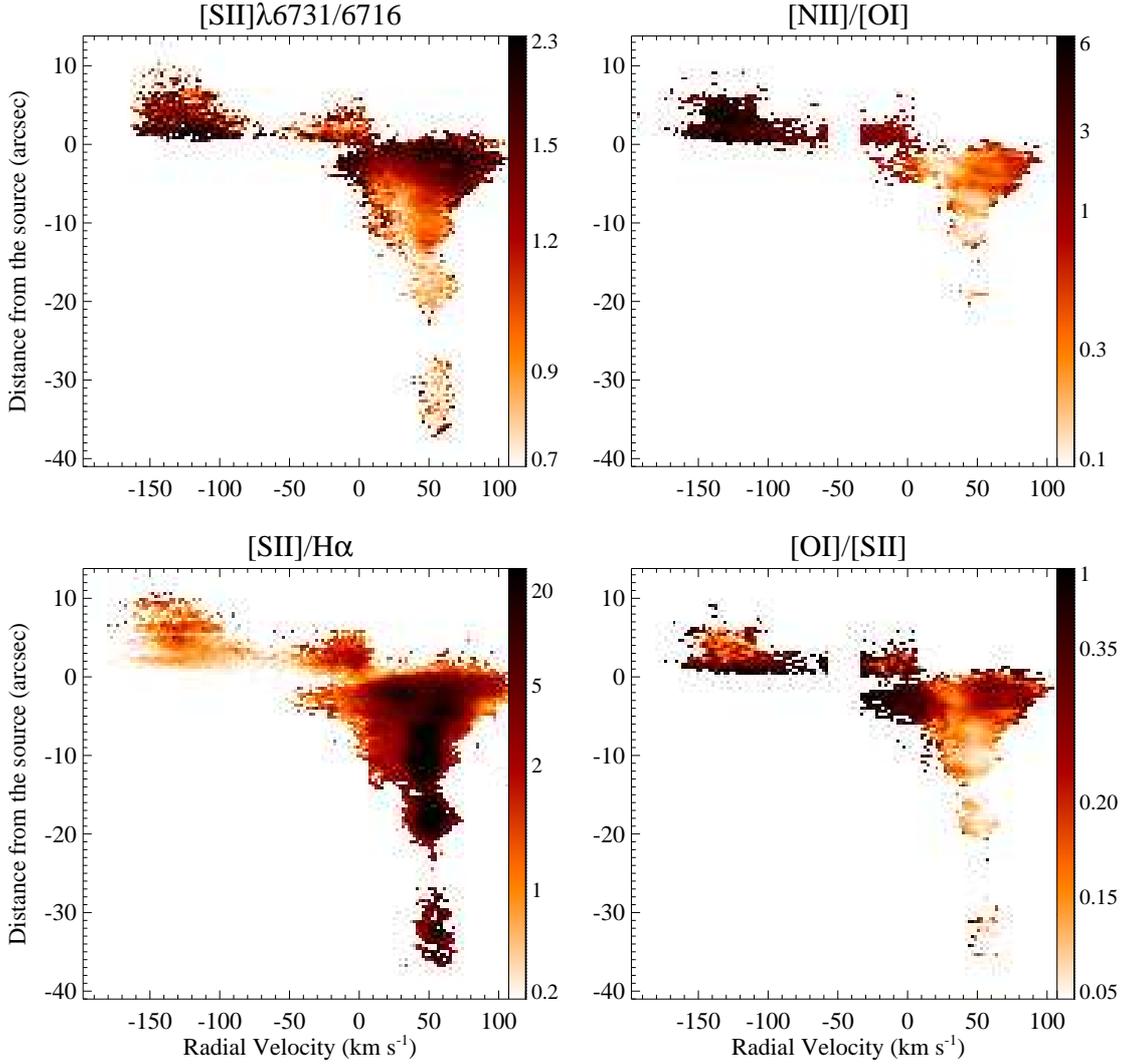}
   \caption{Position-velocity diagrams of the line ratios used to infer the gas physical conditions in the jet:
     the \sii\lam6731/6716 ratio increases with electron density, \en\, ({\it top left});
     the \azii/\oi\, ratio mainly depends on the ionisation fraction, \xe, and
     increases for increasing \xe\, ({\it top right});
     the \sii/\Ha\, ratio is decreasing for increasing excitation conditions ({\it bottom left});
     the \oi/\sii\, ratio depends on all the parameters but is mainly increasing with increasing temperature, 
     \te, and density \en\, ({\it bottom right}).
   To better visualize the variations of the line ratios a
   histogram equivalent color scale is used.}
   \label{fig:ratios}
    \end{figure*}
   \begin{figure*}
     \centering
     \includegraphics[width=15.cm]{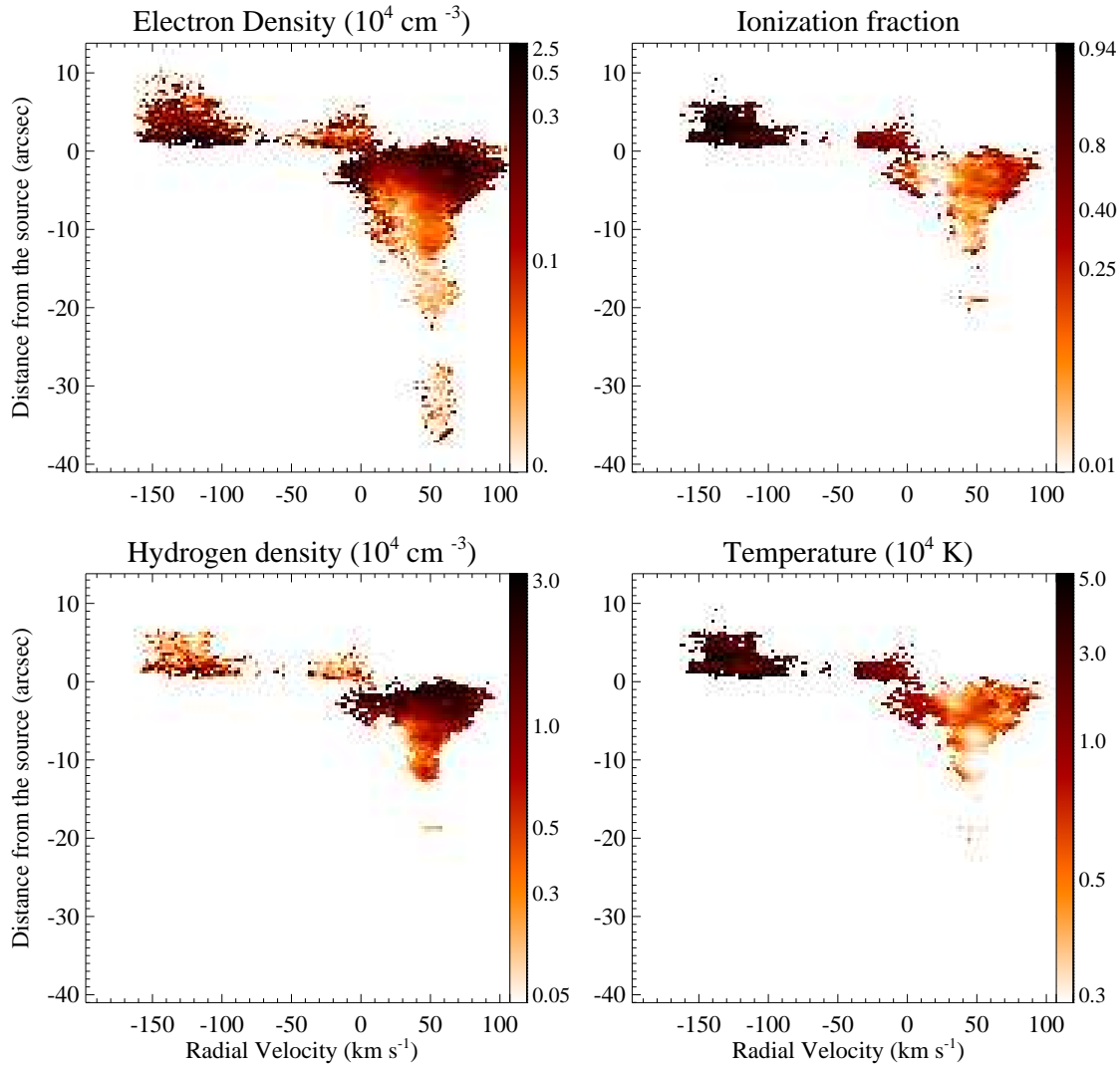}
   \caption{Position-velocity diagrams of the gas physical conditions
     in the jet (a histogram equivalent color scale is used). 
   The electron density, \en, in units of 10$^4$ \cmc\, ({\it top left}), 
   the ionisation fraction, \xe\, ({\it top right}), 
   the temperature, \te, in units of 10$^4$ K ({\it bottom right}), 
   and the total density, \nh, in units of 10$^4$ \cmc\, ({\it bottom left})
   are calculated for each pixel of the spectral images.
    (1 pix=0.378\arcsec\, in the spatial direction and
    2.15~\kms\, in the spectral direction; 
    angular resolution: FWHM$_{seeing}$$\sim$0.8\arcsec$\sim$2 pixels,
    spectral resolution: FWHM$_{inst}$$\sim$8.8~\kms$\sim$4 pixels).
   The errors on the estimated values of \en, \xe, \te, \nh\, may vary between a few percent and 50\%
   depending on the line signal-to-noise on each pixel. 
}
   \label{fig:par}
    \end{figure*}
   \begin{figure*}
     \centering
     \includegraphics[width=15.cm]{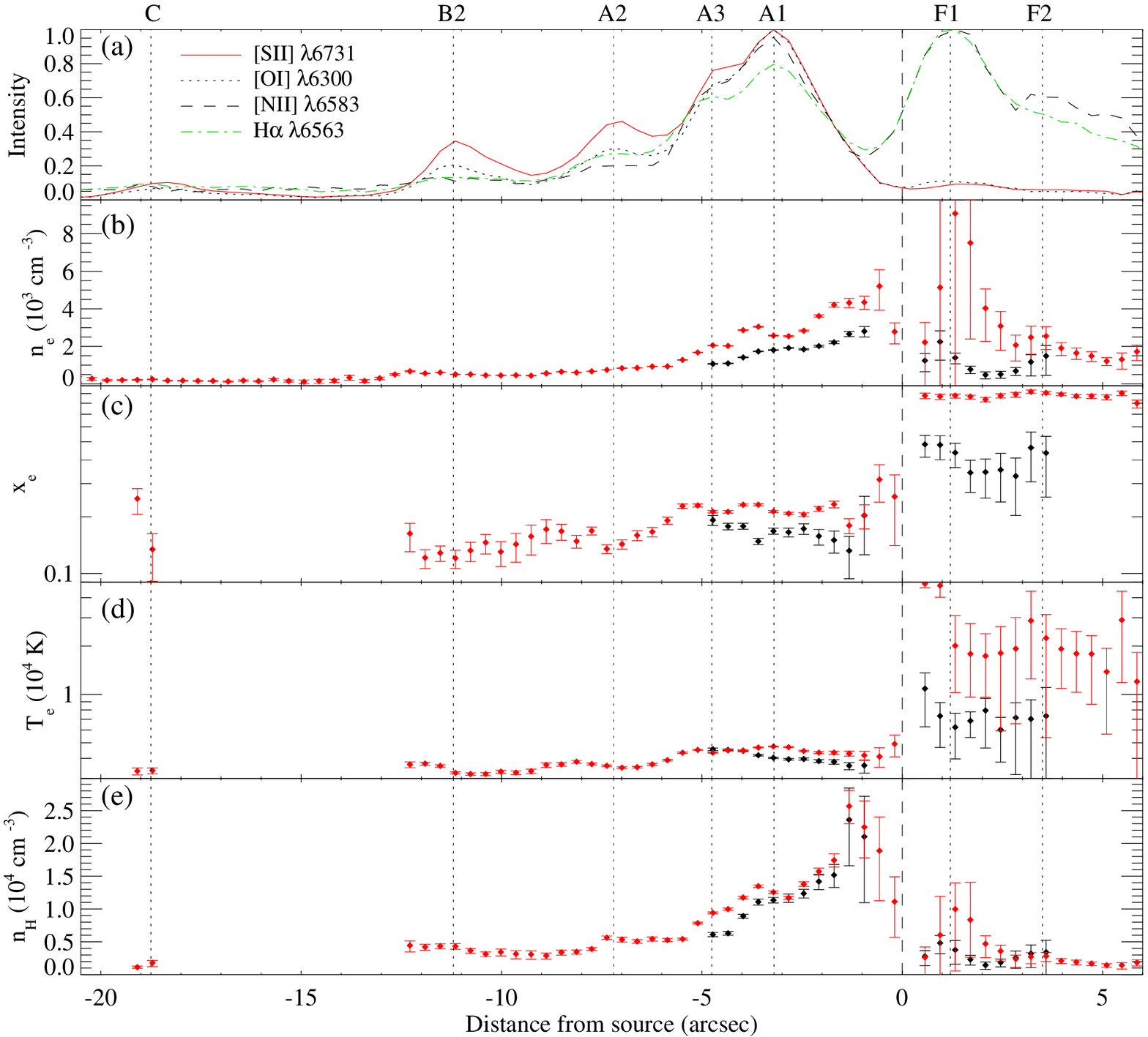}
   \caption{Variations of the gas physical conditions along the jet and
     in the low (LVC, black diamonds) 
     and the high (HVC, red diamonds) velocity components.
     The points are a median of the values shown in Fig.~\ref{fig:par}
     over the HVC and the LVC.
     {\em From top to bottom:} 
     (a) intensity profiles of the \sii, \oi, \azii, and \Ha\, lines integrated over 
     their velocity profile and normalized to their peak intensity
     (b) the electron density, \en, in units of 10$^3$ \cmc;
     (c) the ionisation fraction, \xe;
     (d) the temperature, \te, in units of 10$^4$ K;
     (e) the total density, \nh, in units of 10$^4$ \cmc.}
   \label{fig:phys_med}
    \end{figure*}

The position-velocity diagrams shown in Fig.~\ref{fig:lines_pv} suggest a 
strong asymmetry of the gas excitation conditions in the two jet lobes.
Low excitation lines, such as \sii\lam6731 and \oi\lam6300, are much fainter in the blue lobe than in
the red one, while higher excitation lines, such as \azii\lam6583 and H$\alpha$\,\lam6563, are 
equally bright in the two lobes (see also the top panel of Fig.~\ref{fig:phys_med}).
This suggests that the gas in the blue lobe is not only moving at higher
velocity, as inferred in the previous section,
but is also more excited with respect to the gas in the red lobe.

In order to examine the jet physical conditions we compute selected line ratios which are known 
to be good tracers of the gas excitation  (see Fig.~\ref{fig:ratios}).
\sii\lam6731/6716 probes the  electron density, \en, increasing up 
until \en\, reaches a critical density of 2.5\,10$^4$ \cmc.
On the other hand, \azii\lam(6548+6583)/\oi\lam(6300+6364)  
(hereafter \azii/\oi) and \oi\lam(6300+6364)/\sii\lam(6716+6731) (hereafter \oi/\sii) 
strongly depend on the gas ionisation fraction and temperature \citep{bacciotti99}.
In particular, the \azii/\oi\, line ratio is increasing with \xe\,
while \oi/\sii\, increases with both \te\, and the density.
Finally, the \sii\lam(6716+6731)/\Ha\,\lam6563 line ratio (hereafter \sii/\Ha)
is an indicator of the excitation level
(see, e.g., the predictions from shock models
by \citealt{hartigan94}).
In agreement with \citet{mundt87,mundt91}, we found that \sii/\Ha\, can be as 
high as $\sim$20 in the red lobe and very low ($\sim$0.2-1) in the blue one.
This confirms that the redshifted jet is generally of low excitation, while the
gas in the blue lobe is highly excited.

By applying  the so-called BE technique, described in \citet{bacciotti99},  
to the computed \sii\lam6731/6716,
\azii/\oi, and \oi/\sii\, ratios pixel by pixel we obtain position-velocity 
maps of the parameters characterizing the gas conditions, i.e. 
\en, \xe, \te, and \nh\,  (Fig.~\ref{fig:par}). 
The method relies on the fact that the observed forbidden lines are collisionally excited
and that, in low-excitation conditions, and provided that no strong 
sources of ionising photons are present, the ionisation fraction of hydrogen is
tightly related to those of nitrogen and oxygen via charge exchange processes. 
\citet{bacciotti99} and, later, \citet{nisini05}, verified that the
lines used in the technique are emitted in a region of similar temperature, 
density, and ionisation fraction. 
\citet{hartigan07} further confirmed the validity of the method by applying 
spectral diagnostic techniques to infer the physical properties of the HH 30 jet.
With this technique one can retrieve the gas physical conditions 
comparing the observed line ratios with the ones computed for a grid 
of values of \en, \xe, and \te. 
The errors affecting the estimated parameters are due mainly to the errors on
the measured line fluxes, i.e. on the signal-to-noise (see \citealt{podio06,podio09} for
a discussion on the computation of the errors). 
Note that the diagnostic technique uses ratios between different species and, 
thus, the obtained \xe\, and \te\, may depend on the chosen set of elemental abundances.
However, as stressed in \citet{podio06}, the {\it relative} variations 
of the parameters (i.e. in different knots and/or velocity components) do not depend on the 
choice of an abundance set (see Fig.~1 of \citealt{podio06}) which, therefore,
can be thought of as a model parameter.
Moreover, the most recent determinations by \citet{asplund05} (Solar) 
and \citet{esteban04} (Orion) are in good agreement and produce diagnostic
results that differ by not more than 15\%, which in most cases is smaller
than the uncertainty derived by measurements errors.
Since the inferred physical parameters are used also to determine the
calcium gas-phase abundance with respect to Solar (see Sect.~\ref{sect:ca_depl}), 
we adopt the Solar abundances estimated by \citet{asplund05}, as in \citet{podio09}. 
We did not apply to the spectra a correction for extinction, but we
note that, despite the fact that the source is highly extincted 
by the surrounding optically thick disk 
(A$_{V}$$\sim$6.5 mag or higher, \citealt{jones86,watson04}), 
the Balmer decrement measured in the inner
  knots A and F (H$\alpha$/H$\beta$$\sim$6-7, \citealt{jones86})
  indicate that A$_{V}$ is always $<$1 along the jet.
(According to shock models H$\alpha$/H$\beta$ may vary between
  $\sim$6-7 in low velocity shocks (V$_{s}$$\sim$20~\kms) down to a minimum 
  value of $\sim$3 in very high velocity shocks (V$_{s}$$\sim$100~\kms), \citealt{hartigan94}).
On the other hand, \citet{bacciotti99} show that, since the lines used in this diagnostic 
are very close in wavelength, errors introduced by not correcting 
for extinction are at most 8\%-10\% for the ionization fraction
and 15\% for the temperature, as long as the visual extinction is
lower than $\sim$3. 
To check the obtained \xe\, values we compare the observed \azii\lam(6548+6583)/\azi\lam(5198+5200)
with those computed using the inferred parameters. We selected the pixels with high S/N since
the \azi\, lines are much fainter than the lines used in the diagnostics. For S/N$>$20 we find 
that observed and computed ratios agree within 20\%-30\%, i.e. within the error. 

The PV diagrams in Fig.~\ref{fig:par} show the variation of the electron and total density, \en\,
and \nh, the ionisation fraction, \xe, and the temperature, \te, both along the
two jet lobes and as a function of the gas velocity.
The PV diagrams indicate that the jet is denser and more excited close to the 
source and at high velocities in both lobes.
The diagnostic confirms the strong asymmetry between the two lobes, 
the blue one 
being more excited, with values of the ionisation fraction going
up to $\sim$0.9 and temperatures up to 5 10$^4$ K.

To have a clear view of the variation of the gas physical conditions along the
two jet lobes and in the different velocity components we computed, at each position along the jet, 
the median of \en, \xe, \te, and \nh\, on a $\sim$20 \kms\, velocity interval centered on the peak velocity 
of the low and the high velocity components.
The obtained values are shown in Fig.~\ref{fig:phys_med}, while their average over the first 5\arcsec\, 
where both the HVC and LVC are detected, is reported in Tab.~\ref{tab:avg_par}.
The contribution of the MVC, only barely visible in the blue lobe, is not significant and
it is not reported in the figure and in the table. 
The error bars are the median of the errors on the estimated gas physical parameters 
over the selected $\sim$20 \kms\, interval.
Note that these are much larger in the blue lobe where the signal-to-noise is much
lower.

The electron density \en\, has a symmetrical trend on the first few arcseconds of the
two lobes, decreasing from values of $\sim$6-8 10$^3$ \cmc\, close to the source
to $\sim$1 10$^3$ \cmc\, at $\sim$6\arcsec\, in the HVC,
and from $\sim$3 10$^3$ \cmc\, to $\sim$500 \cmc\, in the LVC. 
It is then even lower in knots C and D of the red lobe, where it 
reaches values of $\sim$200 \cmc.
The excitation conditions, however, are dramatically different in the two lobes. 
The gas in the blue lobe is very hot (\te(HVC)$\sim$1-5 10$^4$ K, \te(LVC)$\sim$0.6-1 10$^4$ K)
and highly ionised (\xe(HVC)$\sim$0.7-0.8 and \xe(LVC)$\sim$0.3-0.5).
On the contrary, in the red lobe the temperature is lower (\te$\sim$3-5 10$^3$ K)
as well as the degree of ionisation (\xe$\sim$0.1-0.3).
Finally, the total hydrogen density is higher in the red lobe, where it decreases from
$\sim$2-2.5 10$^4$ \cmc\, in the first 1.5\arcsec\, down to $\sim$5 10$^3$, $\sim$3 10$^3$,
and $\sim$1 10$^3$  \cmc\, in knots A2, B2 and C, respectively; \nh\,
is much lower in the blue lobe, decreasing from $\sim$1 10$^4$ \cmc\,
in knot F1 to $\sim$1 10$^3$ \cmc\, in knot F2.

The fact that electron and total density peak close to the source, 
where the velocity dispersion is higher, suggests that stronger shocks are occurring
at the jet base which compress and heat the flow.

\section{Discussion}
\label{sect:discussion}

In this section we derive from the inferred jet kinematical and physical properties
crucial information on the dynamics of the outflows, that is 
the mass and momentum transported by the two
jet lobes in their different velocity components. 
We then investigate the depletion of refractory species 
in gas phase, and finally 
we discuss the jet asymmetry and the kinematical/physical
structure of the detected velocity components comparing our results with those
obtained for other jets from T Tauri and younger Class 0/I stars.

\subsection{Jet dynamics: mass and momentum flux rates}
\label{sect:jet_dynamics}

   \begin{figure}
     \centering
     \includegraphics[width=8.5cm]{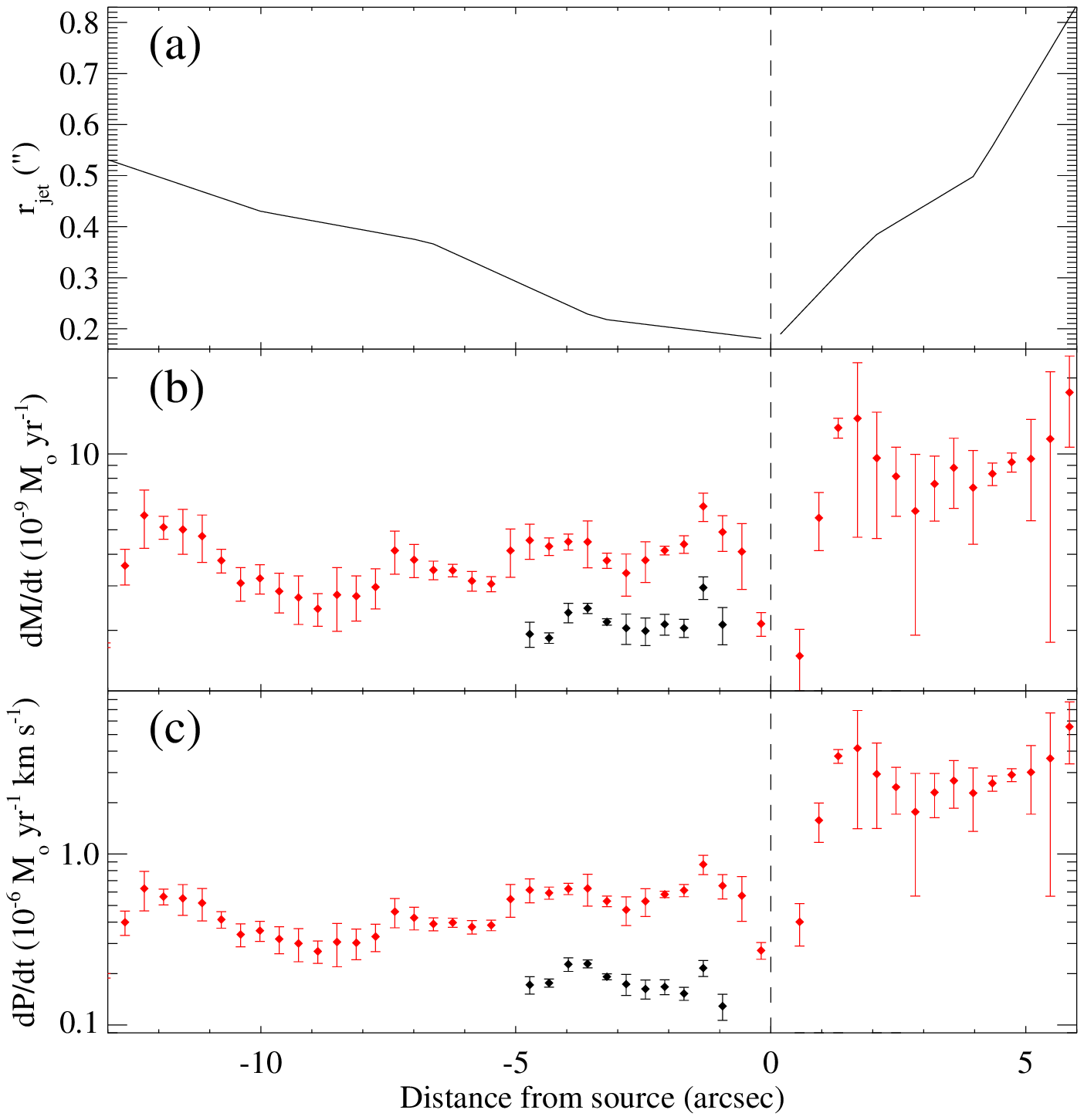}
   \caption{
     {\em From top to bottom:} 
     (a) jet radius, r$_{jet}$, in arcseconds (as estimated by \citet{mundt91} for distances larger than 
     $\sim$2\arcsec\, and $\sim$3.5\arcsec\, in the blue and red lobe,
     respectively, and by \citet{stapelfeldt97}  for lower distances);
     (b, c) mass flux rate, \mjet, in units of 10$^{-9}$ \msolyr, 
     and flux of linear momentum, \pjet, in units of 10$^{-6}$ \msolyr\kms, 
     estimated in the high (HV, red points) and low (LV, black points) velocity components.}
   \label{fig:mdot}
    \end{figure}

The mass and momentum flux transported by the jet, \mjet\, and \pjet,  
are fundamental parameters to understand the   
dynamical relationship between the jet and its environment, and  
to test the validity of the
magneto hydro-dynamical models proposed to explain the jet launch. 
For example, the models predict that the ratio between the rate of mass ejected into the jet 
(\mjet) and accreted from the disk onto the star 
(\macc) is fixed (\mjet / \macc $\sim$ 0.01 - 0.1, \citealt{shu00, konigl00}).

As shown in previous studies \citep{hartigan94,bacciotti99,nisini05,podio06} 
an estimate of \mjet\, can be inferred from the determined total density,  $n_{H}$,
and the measured gas velocity, \vjet,  and jet radius, \rjet, as
\mjet = $\mu$\,$m_{H}$\,$n_{H}$\,$\pi$\,\rjet$^2$\,\vjet,
where $\mu$=1.49 is the mean atomic weight and $m_{H}$ the proton mass.
The flux of linear momentum transported by the jet, \pjet,
is then easily obtained as \pjet = \mjet\, \vjet.
As noted in \citet{podio06} this way of estimating the mass flux 
is independent on reddening and distance estimates.
On the other hand, it assumes that the jet is uniformly filled at the density derived from 
the diagnostics within the estimated jet radius, \rjet, thus 
providing basically an upper limit to \mjet. 
Such an effect is partially compensated for by the presence of regions at higher densities
in the jet than those traced by the \sii\, lines,  which
are not taken into account in the calculations \citep{nisini05,podio06}.

We estimate separately \mjet\, in the low and high velocity components in the two
jet lobes.
The absolute jet velocity, \vjet, is derived from the radial velocities estimated by Gaussian 
fitting of the \sii\lam6731 line profile 
(see Fig.~\ref{fig:kinematics}) and correcting for the inclination angle of 65\degr\, \citep{eisloffel98}.
The jet radius, \rjet, has been estimated by \citet{mundt91} for distances larger than 
$\sim$2\arcsec\, and $\sim$3\farcs5 in the blue and red lobe, respectively, and by \citet{stapelfeldt97} for 
d$\sim$0\farcs1-5\arcsec.
The variation of \rjet\, with distance from the source, obtained by interpolating the estimates from these 
authors, is shown in the upper panel of Fig~\ref{fig:mdot}.
The errors on the inferred values of \mjet\, are obtained by propagating the errors affecting
our estimates of the jet total density, \nh, and of the radial velocity, \vrad.
One should keep in mind, however, that the main source of uncertainty is the jet radius.
For example, \citet{mundt91} estimated \rjet\, by applying one-dimensional deconvolution
method to seeing-limited images of the jet in the \sii\, lines and the uncertainty on the obtained values
can be as high as 30\%.
The values of \rjet\, obtained by \citet{stapelfeldt97} through
HST/WFPC2 broadband imaging are more accurate because the
jet width is resolved thanks to the high angular resolution offered by HST ($\sim$0\farcs1).
In both cases, however, there is no spectral information, thus estimates of \rjet\, relative to the
different velocity components are not available.
Here we assume that the values derived from the works cited above are
valid for all velocities, despite the fact that \citet{bacciotti00}
and \citet{lavalleyfouquet00} showed that, in the case of the jet from DG Tau, 
the high velocity gas is more collimated than the low velocity component.   

The estimated values of \mjet\, in the HVC and in the LVC of the red and the blue lobe 
are plotted in Fig.~\ref{fig:mdot}.   
In the red lobe \mjet\, turns out to be  $\sim$2-6 10$^{-9}$, $\sim$2-3 10$^{-9}$ \msolyr\, 
in the  high and low velocity components, respectively.
In the blue lobe \mjet(HVC)$\sim$0.2-2 10$^{-8}$ \msolyr\, and 
\mjet(LVC)$\sim$0.3-4 10$^{-10}$ \msolyr.


For what concerns the linear momentum, in the red lobe 
\pjet\, turns out to be  $\sim$3-9 10$^{-7}$ and $\sim$1-2 10$^{-7}$
\msolyr\,\kms\, in the high and low velocity components, respectively,
while in the blue lobe it is
$\sim$2-6 10$^{-6}$ and $\sim$0.05-5 10$^{-9}$  \msolyr\,\kms\, in the
HVC and LVC, respectively.
The values of \mjet\, and \pjet\, for the blue lobe LVC are 2-3 orders 
of magnitude lower than in the HVC, because of the very low velocity (only a few \kms) 
coupled to very low total densities (\nh$\le$5\,10$^{3}$ \cmc), and are not reported in Fig.~\ref{fig:mdot}.
In general, the much lower and higher velocities of the LVC and the HVC 
in the blue lobe with respect to the red one, suggest less interaction with the 
ambient medium, as explained in more details in Sect.~\ref{sect:comparison}.

The average of \mjet\, and \pjet\, over the first 5\arcsec\, in the high and 
the low velocity components of both lobes are summarised in Tab.~\ref{tab:avg_par}.   
The  average values of \mjet\, indicate that, summing the HVC and LVC contributions, 
the mass outflow rate transported by the blue lobe (\mjet(blue)$\sim$8.2 10$^{-9}$ \msolyr) 
is only $\sim$1.3 times higher than that transported by the red one 
(\mjet(red)$\sim$6.4 10$^{-9}$ \msolyr). Considering the 
large error on this quantity, it can be said that in practice the amount 
of mass transported by the jet in the two lobes is comparable, despite the large 
differences in density and velocity. This situation, already found for the asymmetric bipolar jet 
from the T Tauri star RW Aur by \citet{melnikov09}, 
indicates that the ejection power is similar on the two sides of the disk,
as it is further discussed in Sect.~\ref{sect:comparison}.

The rate of mass ejected into the jet, \mjet, cannot be compared 
with the mass accretion rate onto the star, \macc,
because DG Tau B is obscured by the optically thick accretion disk, thus
preventing veiling measurements at optical and/or near-infrared wavelengths \citep[e.g., ][]{hartigan95}.
The inferred value of $\sim$6-8 10$^{-9}$ \msolyr\, is  of the
  same order of magnitude of those estimated for more evolved 
T Tauri stars (\mjet(TTSs)$\sim$10$^{-10}$-5\,10$^{-8}$ \msolyr)
through the same method adopted in this paper \citep{coffey08,melnikov09,maurri10} 
or also from the luminosity of optical forbidden lines \citep{hartigan95},
 while it turns out to be lower than the typical values
inferred for HH jets emitted by young Class 0/I sources
(\mjet(Class\,0/I)$\sim$7\,10$^{-8}$-4\,10$^{-7}$ \msolyr, \citealt{hartigan94,bacciotti99,podio06}).
The derived low values of \mjet\, may indicate that the
bulk of the gas is emitting in lower excitation lines lying at
far-infrared wavelengths, such as the \oi\,63\um\, line. This hypothesis seems to be confirmed by
preliminary results from the analysis of Herschel/PACS observations \citep{podio10}.

The linear momentum transported by
the blue lobe (\pjet(blue)$\sim$2.5 10$^{-6}$ \msolyr \kms) is $\sim$3 times higher than 
in the red one (\pjet(red)$\sim$8 10$^{-7}$ \msolyr \kms), due to the multiplication 
by the jet velocity. In order to understand the dynamical relationship between 
the jet and coaxial molecular outflows, it is very interesting 
to compare the estimated value of the momentum flux transported by the jet,
with the momentum flux transported by the surrounding molecular outflow detected in
CO lines. \citet{mitchell94} estimated a momentum flux of $\sim$1.9  10$^{-6}$ \msolyr \kms\, by assuming
a distance of $\sim$160 pc and an inclination angle with respect to the sky of $\sim$15\degr.
Correcting this value for recent more accurate estimates of the
distance (d$\sim$140 pc) and of the  
inclination angle (i$\sim$25\degr, \citealt{eisloffel98}) one obtains 
\poutflow$\sim$10$^{-6}$ \msolyr \kms.
This value is comparable to the momentum estimated for the jet,
indicating that the latter is powerful enough to drive the molecular outflow detected
at millimeter wavelengths.
How precisely the jet transfers its linear momentum to the surrounding
medium, however, remains an open question (see, e.g., \citealt{downes07}).

\subsection{Dust reprocessing along the jet: investigating Ca, Ni, and Fe gas-phase abundance}
\label{sect:ca_depl}

   \begin{figure}
     \centering
     \includegraphics[width=7.cm]{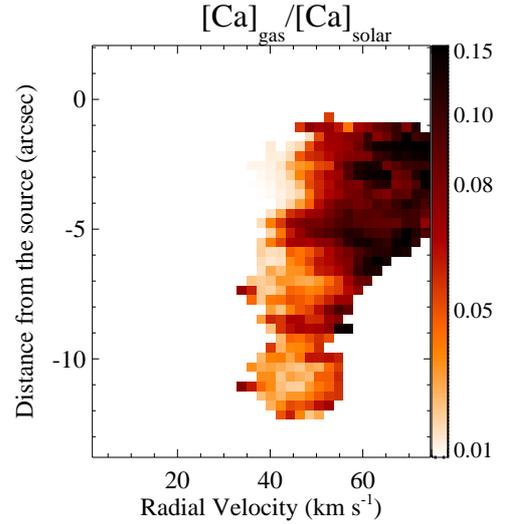}
   \caption{Position-velocity diagrams of Ca gas-phase abundance estimated as 
     the ratio between observed and predicted (by assuming Ca solar abundance) 
     \caii\lam7291/\sii\lam6731 ratios. 
   There is more calcium in gas phase at
   high velocities, similarly to what found in \citet{podio09}, and in
   agreement with models predictions of dust destruction in shocks \citep{guillet09}.}
   \label{fig:ca}
    \end{figure}

The KECK spectra of DG Tau B include also a number of emission lines from ``refractory'' species, such as
Fe, Ni, and Ca.
These species are known to be highly depleted in the interstellar medium because their atoms
may be locked onto dust grains \citep{savage96}.
Estimates of Ca, Fe, and Ni gas-phase abundance in Orion and $\zeta$ Oph 
indicate that they are depleted up to four orders of magnitude with respect to their solar abundances
\citep{baldwin91,savage96,esteban04}.
On the other hand, dust grains can be partially or completely
destroyed both in the jet launching region and in the shocks occurring
along the flow, releasing Ca, Fe, and Ni atoms into the gas-phase.
 
Both the models \citep{isella05} and near-infrared interferometric observations 
\citep{akeson05,eisner07} show that the inner part of the accretion disk around T Tauri and 
Herbig Ae/Be stars is dust-free, 
because dust grains are destroyed by the stellar radiation up to the
so-called dust evaporation radius, R$_{evp}$. This is located at $\sim$0.1-1.5 AU in Herbig stars 
(Eisner et al. 2007) and at $\sim$0.05-0.3 AU in T Tauri stars (Akeson et al. 2005).
For younger Class I sources there are no direct estimates of R$_{evp}$ because of their
embedded nature.
However, we can obtain a rough estimate of the position of R$_{evp}$ for DG Tau B, assuming that 
the model by \citet{isella05} is still valid for such a young source.
For a luminosity of $\sim$0.7-0.9 L$_{\odot}$ \citep{jones86,watson04} and a 
mass accretion rate of at least $\sim$10 times higher the inferred \mjet\, we 
find that R$_{evp}$ is located at about 1-6 times the position of the
truncation radius, R$_{t}$, where  the stellar magnetosphere truncates the disk.
(The position of R$_{t}$ is computed from equation (2.2a,b) of \citet{shu94}, assuming
typical temperature and mass for DG Tau B spectral type ($\le$K6,
\citealt{watson04}), T$\sim$3000-5000 K and M$\sim$0.1-1 M$_{\odot}$, and
stellar magnetic field strength, B$_{*}$, of $\sim$0.5-3 kG, \citealt{johns-krull07}). 
Thus, if the jet originates in the inner region of the disk ($r$$<$R$_{evp}$) we expect
Ca, Fe, and Ni to be present in solar abundance.
On the contrary, if the jet originates from a region of the disk
extending beyond  R$_{evp}$, part of the dust is lifted with the jet
material, and these species will be observed in the jet to be depleted with respect to their solar abundances.

Also the shocks occurring along the jet can affect the gas-phase abundances of refractory species.  
Theoretical models of C- and J-shocks showed that processes such as  thermal 
and inertial sputtering, photoevaporation and shattering, can destroy from a few percent up to
$\sim$90\% of the dust grains depending on a number of parameters such as the shock velocity, 
the gas pre-shock density, the intensity of the magnetic field, and the size and 
structure of the dust grains 
\citep[e.g., ][]{jones94,jones00,may00,draine03,guillet09}. 

Thus, investigating the gas-phase abundances of Ca, Fe, and Ni along jets we can test observationally
the predictions of jet launching models and dust reprocessing in shocks.

To this aim we analyse the spatio-kinematical distribution of the
emission from Ca, Fe, and Ni in the jet, and we estimate the
gas-phase abundance of calcium, [Ca/H]$_{gas}$ (hereafter [Ca]$_{gas}$).
The PV diagrams in Fig.~\ref{fig:faint_lines_pv} show that the
emission in these lines is confined to the high velocities both in the red and in the blue lobe.
The velocity profiles suggest that the atoms of Ca, Fe, and Ni, highly depleted in the ISM, may be
released in gas-phase in the inner disk region and/or in the shocks.
In both cases higher abundances are expected in the high velocity component.  
In fact, on the one hand, the jet launch models predict that the gas  
extracted from the inner disk region is accelerated to
higher velocities with respect to the gas extracted from the outer disk.
On the other hand,
the destruction rate in a shock is highly dependent on the shock velocity.
Thus higher destruction rate is expected in the strongest shocks and at high velocities.

We estimate the Ca gas-phase abundance following the procedure illustrated 
in \citet{nisini05} and \citet{podio06,podio09}, that
consists of comparing observed and predicted ratios between refractory and non-refractory 
species, such as sulphur, 
which are expected to be present in solar abundance in the interstellar medium.
Following these studies we compare observed
\caii\lam7291/\sii\lam6731 ratios (hereafter \caii/\sii), with those computed through the estimated parameters (\en, \xe, and \te) and assuming
solar abundance for both Ca and S \citep{asplund05}.
The abundance of calcium in gas-phase in the jet with respect to its solar
value is given by the ratio between
observed and predicted   \caii/\sii\, ratios: 
[Ca]$_{gas}$/[Ca]$_{solar}$=(\caii/\sii)$_{obs}$/(\caii/\sii)$_{predicted}$.
Following \citet{hartigan04} we assume that the \caii\, lines are collisionally excited and compute the 
level populations. 
As in previous works \citep{nisini05,podio06,podio09} we also assume that there is no calcium in the form of 
Ca$^0$, since its ionisation potential is very low, $\sim$6.1\,eV.
(Note that emission in the \nai\, D lines is barely detected only at the
  peak velocity of knots A1, A3 at $\sim$6 $\sigma$).
Also we assume that at every position along 
the jet, the amount of calcium ionised twice with respect to singly ionised calcium, i.e. Ca$^{++}$/Ca$^{+}$,
is equal to the hydrogen ionisation fraction, \xe.
As explained in \citet{podio09} the latter assumption is justified by
the fact that the ionisation potential of Ca$^{+}$ is similar to that of hydrogen  
($\sim$11.9\,eV and $\sim$13.6\,eV, respectively), as are the recombination and collisional ionisation
coefficients for temperatures lower than 3\,10$^4$ K. Thus, Ca$^{+}$ and H$^{0}$ atoms can be thought of 
as undergoing the same processes when moving along the jet.

We have no estimates of the visual extinction, thus we cannot correct the Ca$^{+}$ and S$^{+}$  line 
fluxes for reddening. 
This implies that the observed ratios (assuming  A$_{\rm V}$=0) are actually upper limits 
to the real values (which decrease for increasing A$_{\rm V}$). 
Thus, the calcium depletion inferred from our analysis is a lower limit. 
 
Even so, the results shown in Fig.~\ref{fig:ca} show that
Calcium turns out to be strongly depleted, its gas-phase abundance varying between a few percent and 15\% 
of its solar value ([Ca/H]$_{solar}$$\sim$2.04 10$^{-6}$, \citealt{asplund05}).
Interestingly, the Ca abundance is at a maximum close to the source and at high velocities, where the 
broad velocity profiles and high values of \en\, and \xe\, suggest the presence of strong shocks.
We also derived an estimate of the Ca gas-phase abundance in the blue lobe, despite the
very low signal-to-noise of calcium emission here.
It turns out that the observed \caii/\sii\, ratio is $\sim$1-3
times the predicted one, implying that Ca is completely
released into the gas-phase in the blue lobe, where the very high excitation conditions
(\xe, \te) indicate very high shock velocities. 
Similar results have also been found in the HH 111 protostellar jet
\citep{podio09} and confirm the dependence of elemental depletion on shock
 velocity as predicted by shock dust destruction models \citep[e.g., ][]{guillet09}.

\subsection{Investigating the origin of velocity components and
  asymmetries in stellar jets}
\label{sect:comparison}

The presence of a low and a high velocity components in jets from
young stars was first noted by \citet{hamann94}. 
\citet{hirth97} analysed \sii, \oi, and \azii\, forbidden emission lines in 38 TTSs and
report 58\% of them showing a LVC and a HVC, with the LVC being 
always less extended than the HVC. 
Similarly, two velocity components have been detected at the base of HH jets emitted by 
Class 0/I sources and, also in these cases, the LVC is  less
  extended than the HVC. However, while in the flow from T Tauri
  stars the LVC is confined within $\sim$20-200 AU \citep{hirth97},
  the jets emitted by younger Class 0/I sources present LVC extending
  up to $\sim$1000-2000 AU from the source
  \citep{garcialopez08,garcialopez10}.
In agreement with these results the LVC from DG Tau B is observed up
to $\sim$500-700 AU from the source.

Moreover, \citet{hirth97} note that the LVC is not observed in the \azii\, lines, suggesting that the LVC
is less excited than the HVC, i.e. N is ionised only once in the low velocity gas.
More recent deeper observations, however, have shown low velocity \azii\, emission in a few
objects, such as DG Tau and CW Tau, allowing us to apply spectral diagnostic techniques and to
infer the physical properties of the gas in the two velocity components. 
Therefore, we compare our results with those found for DG Tau and CW Tau to
investigate the properties and the origin of the high and the low velocity components in jets.

The values of the physical parameters summarised in Tab.~\ref{tab:avg_par} show that in the jet
from DG Tau B the HVC is denser and more excited than the LVC in both the blue and the red lobe.
Higher electron densities and ionisation fractions in the HVC have been found also in DG Tau
\citep{lavalleyfouquet00,maurri10}, and in CW Tau \citep{coffey08}.
On the other hand, the analysis of the line ratios in a large sample
of TTSs by \citet{hamann94} suggest higher temperatures but
lower electron densities in the HVC.
Combining  the results from DG Tau B with these previous
estimates, we can conclude that there is 
clear evidence that the HVC is more excited than the LVC. 
To understand if the HVC is also denser is more difficult since in many cases an accurate estimate of the 
total density is limited by the fact that the \sii\, lines used to infer \en\, are saturated in one
or both velocity components, hence obtaining only lower limits to the real density
\citep{lavalleyfouquet00,coffey08,maurri10}. 
For jet from Class 0/I sources, the electron density, estimated
through near-infrared \feii\, lines, is higher at high velocities
along HH 1 and at low velocities in HH 34 and HH 46-47. 
Since no estimate of the ionisation fraction are available, however, it is
not possible to understand which is the denser component.
Interestingly, higher densities and excitation conditions at high velocities have been noted
at the position of the shock surfaces detected along the HH 111 jet 20-60\arcsec\, away from the source 
\citep{podio09}.


The fact that the LVC shows lower excitation conditions suggests that this component
is originated in an extended disk-wind where the gas extracted from an outer region of the disk with respect
to that emitting the HVC is accelerated at lower velocities and give rise to slower shocks (hence, the 
lower ionisation fraction and temperature).
This hypothesis seems to be confirmed by high angular resolution
observations of DG Tau obtained 
with adaptive optics system and/or with the Hubble Space Telescope
which resolve the jet perpendicular
to its axis and show that the HV gas is more collimated than the LV gas 
\citep{bacciotti00,lavalleyfouquet00,maurri10}. 


Another interesting and still unexplained characteristic of jets from
young stars is that they are often asymmetric.
\citet{hirth94} noted that 8 out of 15 objects  for which estimates of the radial velocities in both lobes
are available show a velocity asymmetry with the ratio between the velocity of the fast lobe and the 
velocity of the slow lobe, V$_{f}$/V$_{s}$, ranging between 1.4 and 2.6.
However, a statistical study of asymmetric jets, aimed to investigate the
origin of such an asymmetry, is hampered by the fact that in most of 
the cases one lobe is much more extincted or intrinsically fainter
than the other one making it difficult to derive the gas properties in both lobes.
Our deep KECK/HIRES observations allows for the first time to detect
both lobes of DG Tau B in all the emission lines which are employed in our
diagnostic techniques and thus to derive \en, \xe, \te, and \nh\,
in both lobes.

We compared our results against those found by \citet{melnikov09} for
the well-known asymmetric jet from RW Aur,
finding  that the jets from the two stars have very similar
kinematical and physical characteristics.
First of all, in both cases 
the blue lobes are unusually much fainter than the red ones.
We note that in both cases the density of the blue lobe is lower, implying 
intrinsically fainter emission. The blue lobe is also faster in both 
targets, probably indicating less interaction with the environment on that 
side of the system. In more detail, 
the blue lobe is $\sim$1.8 times 
faster than the red one in RW Aur and $\sim$2.3 in DG Tau B, 
it is more ionised by a factor of $\sim$3 in RW Aur and $\sim$3.3 in DG Tau B and in both jets 
the fast lobe is less dense than the slow one by a factor  $\sim$0.3. 
An important difference between the two jets is that while the width
of the jet from RW Aur is similar 
in the two lobes (\rjet\,$_{f}$/\rjet\,$_{s}$$\sim$1),  
the available images of DG Tau B indicate that the
fast lobe is less collimated than the slow one (\rjet\,$_{f}$/\rjet\,$_{s}$$\sim$2). 
The most interesting similarity, however, is that in both DG Tau B and 
RW Aur the mass transported by the two lobes is comparable 
(with a ratio of about 1.3 in both targets, smaller than the error on the estimates) 
despite the observed asymmetry in the physical quantities. 
Finally,  the flux of linear momentum transported by the
fast lobe is obviously larger than that transported by the slow one in both systems
(\pjet\,$_{f}$/\pjet\,$_{s}$$\sim$1.4 for RW Aur and $\sim$3 for DG Tau B).
We note that similar \mjet\, in the two jet lobes  has also been found in other stellar jets
showing asymmetries in the physical conditions as, e.g., the HH~30 jet  
\citep{bacciotti99b} and the bipolar jet from the Herbig Ae/Be star LkH$\alpha$~233  
\citep{melnikov08}. 

The striking similarity between the properties of 
these asymmetric jets suggests that similar effects are at work.
As discussed in \citet{melnikov09}, the fact that the global mass outflow rate is 
comparable on the two sides of the DG Tau B system  
appears to indicate that, as much as in the RW Aur case, the power transferred 
by the rotating disk is the same on both sides of the disk, supporting the validity 
of a magneto-centrifugal mechanism. 
In this scenario, however, an asymmetric ambient medium can lead to an asymmetric 
interaction of the disk with the environment, hence different mass load,  lever arms and/or radial extensions 
of the disk region producing the jet on the two sides
\citep{ferreira06}. For example, 
if the ambient radiation field is stronger on one side, 
this increases the level
of ionisation on the surface of the disk  on that side, leading to enhanced mass load on 
the magnetic field lines, and/or to a larger jet launching region. 
In the case of DG Tau B, 
larger mass-load and interaction with the surrounding medium in the red lobe is suggested 
by the detection of a redshifted molecular outflow in the millimeter CO lines \citep{mitchell94}.


\section{Summary and Conclusions}
\label{sect:conclusions}

In this paper we investigate the kinematical and physical structure of
the HH 159 jet emitted by the young Class I source DG Tau B, by means of high
spectral resolution observations acquired with KECK/HIRES.
With respect to previous studies (e.g., \citealt{eisloffel98}) we
couple the analysis of the jet kinematics with a detailed study of its
physical and dynamical properties in each of its lobes and velocity components.
Our method involves the analysis of selected line ratios through the so-called BE diagnostic technique
\citep{bacciotti99} to infer the gas physical conditions (\en, \xe, \te, and \nh).
Our main conclusions are as follows:

\begin{itemize}
\item[-] the velocity variations along the jet indicates that the ejection direction is changing 
due to precession and/or interaction with the surrounding medium.
The electron and total density, as well as the ionisation fraction and the temperature, are
decreasing along both lobes, due to    
gas dilution, i.e. to the jet propagation in a conical geometry, and/or to stronger
shocks at the jet base, where the ejected material interacts with the dense parental cloud. \\

\item[-] the lines show a complex velocity and excitation structure, with multiple
velocity components. The high velocity component, HVC, is 
extending more than the low velocity component, LVC, which is fading at 500-700 AU from the source, 
similarly to what previously observed for other jets from Class 0/I sources \citep{garcialopez08}
and, on smaller scales, for jets from CTTSs \citep[e.g., ][]{hirth97}. 
The HVC appears to be denser and more excited suggesting 
that the LV gas is extracted from an outer region of the disk with respect to the HV
gas and is thus accelerated at lower velocities.\\

\item[-] the jet show a strong asymmetry:
the blue lobe is faster and more excited than the red lobe but less dense and collimated, suggesting 
that the interaction of the ejected material with the ambient medium is stronger on the red-lobe 
side. This asymmetry is similar to the one observed in other jets, and in particular 
in RW Aur \citep{melnikov09}.\\

\item[-] despite the observed asymmetries, the mass loss rate is similar in the two lobes
(\mjet$\sim$6-8 10$^{-9}$ \msolyr), as in other asymmetric jets.
This means that the the power transferred by the rotating disk is the same on both sides of the
system, as expected if the jet is launched by a magneto-centrifugal engine.
The observed asymmetries can be explained  in the framework of a MHD disk wind,
if the latter propagates in an inhomogeneous ambient medium \citep{ferreira06}.\\

\item[-] the redshifted molecular outflow detected in the CO millimeter lines \citep{mitchell94} 
supports the idea of an asymmetric ambient medium and a larger mass-load on the red-lobe side. 
The flux of linear momentum transported by the jet is comparable to that estimated
for the molecular outflow, thus the latter can be jet-driven 
(\pjet$\sim$0.8-2.5 10$^{-6}$ \msolyr \kms).\\

\item[-] the depletion of Ca gas-phase abundance with respect to its solar abundance
([Ca]$_{gas}$/[Ca]$_{solar}$$\sim$0.01-0.15) indicates that most Ca atoms are locked 
onto dust grains. The minimum depletion is observed close to the source and
in the HVC, where the strongest shocks are occurring, in agreement with the predictions from
models of dust reprocessing in shocks \citep[e.g., ][]{guillet09}. 
The presence of dust in the jet may imply that the material is extracted from a 
region of the disk extending beyond the so-called dust evaporation radius, R$_{evp}$.
However, more stringent constraints on the jet launching region can 
can only be derived by analysing higher angular resolution observations, to infer Ca gas-phase abundance 
in the first tens of AU from the source where the gas has not been reprocessed by shocks, 
and for more unembedded sources, for which the location of 
R$_{evp}$ can be derived from near-infrared interferometric observations \citep[e.g., ][]{akeson05}. 

\end{itemize}

\begin{acknowledgements}
The data presented herein were obtained at the W. M. Keck Observatory,
which is operated as a scientific partnership among the
California Institute of Technology, the University of California,
and the National Aeronautics and Space Administration.
The Observatory was made possible by the generous
financial support of the W. M. Keck Foundation.
The authors wish to recognize and acknowledge the very significant cultural 
role and reverence that the summit of Mauna Kea has always had within the 
indigenous Hawaiian community.  We are most fortunate to have the opportunity 
to conduct observations from this mountain.
J.E. thanks the Institute for Astronomy, UH, for its hospitality
during his stay as an Otto-Hahn fellow of the Max-Planck Society.
L.P. thanks the Irish Research Council for Science, Engineering and 
      Technology which partially funded her work. 
Finally, this work was partially 
      supported by the European Community's Marie Curie Research and Training 
      Network JETSET (Jet Simulations, Experiments and Theory) under contract 
      MRTN-CT-2004-005592.
\end{acknowledgements}

\input{referenc}

\newpage

\end{document}

%% file: referenc.tex
%
%
%